\documentclass[epj]{svjour}
\usepackage{graphics}
\begin{document}
\title{Calculating broad neutron resonances in a cut-off Woods-Saxon potential}
\author{\'A. Baran\inst{1} \and Cs. Nosz\'aly\inst{1} \and P. Salamon\inst{2} \and T. Vertse\inst{1,2}}
%
\offprints{}          
\institute{University of Debrecen, Faculty of Informatics, PO Box 12, H--4010 Debrecen, Hungary \and Institute for Nuclear Research Hungarian Academy of Sciences (ATOMKI), Debrecen, PO Box 51, H--4001, Hungary}
\date{Received: date / Revised version: date}
%
\abstract{
In a cut-off Woods-Saxon (CWS) potential with realistic depth $S$-matrix poles being far from the imaginary wave number axis form a sequence where the distances of the consecutive resonances are inversely proportional
with the cut-off radius value, which is an unphysical parameter. Other poles lying closer to the imaginary wave number axis might have
trajectories with  irregular shapes  as the depth of the potential increases. Poles being close repel each other, and their repulsion is responsible for the
changes of the directions of the corresponding trajectories. 
The repulsion might cause that certain resonances  become antibound and later resonances again when they collide on the imaginary axis. 
The interaction is extremely sensitive to the cut-off radius value, which is an apparent handicap of the CWS potential. 
} 
\maketitle

\section{Introduction}
Gamow shell model (GSM) \cite{[Mi09]} became a useful tool in analyzing drip line nuclei produced
in laboratories with radioactive beam facilities. A most recent analysis of this type is  that of the $^7$Be(p,$\gamma$)$^8$B and the $^7$Li(n,$\gamma$)$^8$Li reactions in Ref. \cite{[Fo15]}.
The key elements of GSM are the Berggren-ensembles of single particle
states. The   Berggren-ensemble might include resonant and sometime antibound
states beside the bound states and the scattering states along a complex path.
The shape of the path determines the set of the $S$-matrix pole states to be included in the
ensemble. In order to have smooth contribution from the scattering states the shape of the path should go reasonably far from the poles. Therefore
to know, where the poles are located has crucial importance.

The most common phenomenological nuclear potential used is the cut-off Woods-Saxon (CWS) potential. Two handicaps of the CWS potential are that the positions of the broad resonances do depend on the cut-off radius and the pole trajectories also
strongly  depend on this radius, which is
an unphysical parameter. Moreover some of the trajectories show strange circulating shapes \cite{[Ve12]}. Two of us suggested an alternative
potential, the SV potential \cite{[Sa08]} which goes to zero smoothly at a finite distance. With SV
potential the strange circulating shapes are missing, therefore it was conjectured \cite{[Ve12]} that the circulation is due to the jump of the CWS at the cut-off radius.
In spite of of these handicaps the popularity of using CWS potential is still
survives, therefore in this work we study further the pole distribution and the
pole trajectories in CWS potential.

\section{Formalism}

Very little is known about the  positions of the
broad resonances which correspond to poles lying farther from the real energy
or wave-number axes.
By varying the strength of the potential we can calculate pole trajectories on the complex wave number ($k$) plane. The decaying resonances lie in the fourth quadrant of the $k$ plane.  The capturing resonances
are mirror images of the decaying resonances in a real potential, therefore their trajectories are
simply reflected to the imaginary $k$-axis.
 The complex wave number of a Gamow resonance is $k=k^R+ i k^I$ with $k^I<0$, and $k^R>0$  for a decaying resonance, and $k^R<0$ for a capturing resonance. The energy is proportional to $k^2$, therefore the unbound poles lie on the second  energy-sheet. Resonances lying close to the imaginary $k$-axis, when
 $-\pi<arg(k)< -\pi/2$, i.e. $|k^R|<|k^I|$, are called as sub-threshold resonances \cite{[Mu10]}.
 We want to calculate poles of the  $S$-matrix in a given partial wave
having orbital angular momentum $l$.
The partial wave solution satisfies the radial Schroedinger equation
\begin{equation}
\label{radsch}
u^{\prime\prime}(r,k)+k_l^2(r) u(r,k)=0~,
\end{equation}
where prime denotes the derivative with respect to (wrt) the radial distance $r$. The function
\begin{equation}
\label{locwn}
k_l^2(r)=[k^2 -\frac{l(l+1)}{r^2}-v(r)]
\end{equation}
is the so called {\it squared local wave number}.
 
In this work we consider a very simple case in which a spin-less neutral particle
is scattered on a spherically symmetric target nucleus represented
by a  real CWS type nuclear potential $v(r)$.
 The first boundary condition (BC) for the solution $u(r,k)$ is its regularity at $r=0$: 
\begin{equation}
\label{regular}
u(0,k)=0~.
\end{equation}
The other BC is specified
at large distance in the asymptotic region, at $R_{ass}$, i.e.  beyond the cut-off radius $R_{max}$, where
\begin{equation}
\label{sumpt}
v(R_{ass}\ge R_{max})=0~.
\end{equation}
 For a scattering state the asymptotic BC requires that the solution $u(r,k)$
 should be a combination of the incoming $H_l^-(kr)$ and outgoing $H_l^+(kr)$ 
free waves satisfying the Ricatti-Hankel differential equation:
 \begin{equation}
\label{scattbc}
u(r,k)=A [H_l^-(kr)-S(k)H_l^+(kr)]~.
\end{equation}
Here $S(k)$ is the element of the scattering matrix in the $l$-th partial wave. 

For the CWS potential  the  radial equation in Eq. (\ref{radsch}) can be solved only numerically.
At $r=R_{ass}$ the numerical solution should match to that of the asymptotic equation in Eq. (\ref{scattbc}).
We can calculate $S(k)$ from the logarithmic derivative at $R_{ass}$:
\begin{equation}
\label{lgder}
z_i(R_{ass},k)=\frac{u_i^\prime (R_{ass},k)}{u_i(R_{ass},k)} ~,
\end{equation}
were $u_i(r,k)$ is the internal solution, which is regular at $r=0$.

The value of the $S$-matrix at a given wave number $k$ can be calculated as:
\begin{equation}
\label{smatr}
S(k)=\frac{k {\bar H}^-_l(kR_{ass})-z_i(R_{ass},k) H^-_l(kR_{ass})}{k{\bar H}^+_l(kR_{ass})-z_i(R_{ass},k) H^+_l(kR_{ass})}~,
\end{equation}
where bar denotes derivative wrt the argument $kr$.

\section{Resonances as poles of the $S$-matrix}
 For a resonance the asymptotic BC is defined differently. It is required that the solution $u(r,k)$ and  its
 derivative $u^\prime(r,k)$ should be outgoing type and  the logarithmic derivative of the external solution is
\begin{equation}
\label{resonz}
z_e(r,k)=k \frac {{\bar H}^+_l(kr)}{H^+_l(kr)}~.
\end{equation}
Both BC (in Eqs. (\ref{regular}) and  (\ref{resonz})) can be satisfied simultaneously only at discrete complex $k$ eigenvalues, at the  poles of $S(k)$. Here the complex $k$ eigenvalue  is fixed by the zeros of the difference of the logarithmic derivatives of the internal and the external solutions 
\begin{equation}
\label{logder}
G(k,r)=z_i(r,k)-z_e(r,k)~.
\end{equation}
The computer programs GAMOW \cite{[Ve82]}, and ANTI \cite{[Ix95]}  find the zeros of $G(k,r)$, at certain $R_m$ matching radius $0<R_m<R_{ass}$.
 For a broad resonance the proper choice of this $R_m$ is difficult. The zero  is searched by Newton iterations, and the iteration
process often converges poorly or fails. Therefore we developed a new method in which we compare the logarithmic derivatives in a wide region in $r$. This method is built into the program JOZSO\footnote{The program name is chosen to honor   the late J\'ozsef Zim\'anyi to whom one of the authors (T. Vertse) is grateful for starting his carrier.} \cite {[No15]}.
  We calculate $G(k,r)$ in Eq. (\ref{logder})  at equidistant mesh-points with mesh size $h$
 at $r_j=j h$, $j\in [i_1,i_2]$.  The  mesh points are taken from a region where the nuclear potential falls most rapidly.
Then
we search for the absolute minimum of the function of two real variables 
$k^R$, and $k^I$:
 \begin{equation}
\label{minima1}
F(k^R,k^I)=\log [\sum_{j=i_1}^{i_2}  \large{|} G(k,r_j)\large{|]}~.
\end{equation}
 Absolute minima of the function ${F}(k^R,k^I)$ in Eq. (\ref{minima1})
should have a large negative value. The position of the absolute minimum is  the pole position of $S(k)$. The  minimum of the function is found by using the Powell's method in Ref. \cite{[numrec]}. 
  To find the minima
of the  function $F(k^R,k^I)$  first we explore the landscape of the
function $F(k)$  in a complex $k$ domain of our interest.
We plot the function $-F(k)$, which has peaks at the poles of $S(k)$ on a grid
(see e.g. Fig. \ref{modws15}).

\section{Pole trajectories in CWS potential}
The nuclear potential in Eq. (\ref{locwn}) is in fm$^{-2}$ units, like the $k^2$ is.
It is more usual however, to give it the same units in which the energy is given, i.e. in MeV. It is also usual to give the
 potential as a product of its strength $V_0$ in MeV and its dimensionless radial shape:
\begin{equation}
\label{WSpot}
V^{\rm CWS}(r,R,a,R_{\rm max})=-V_0f_{\rm CWS}(r,R,a,R_{\rm max})~,
\end{equation}
where the later is
\begin{equation}
\label{vagottWS}
f_{\rm CWS}(r,R,a,R_{\rm max})=\theta(R_{\rm max}-r)
\frac{1}{1+e^{\frac{r-R}{a}}}~.
\end{equation}
Here $\theta(x)$ denotes the 
Heaviside step function, being zero for negative
and unity for non-negative arguments.
The parameters of the shape in Eq. (\ref{vagottWS}) are the radius $R$, the diffuseness $a$ and the
cut-off radius $R_{\rm max}$.

Sometimes it is useful to draw the trajectories of the individual poles, i.e. to explore how the positions of the poles change when
we keep the shape of the CWS potential fixed but vary the $V_0$ strength of the potential.
 By increasing $V_0$ the pole  trajectory moves to the upper half of the $k$-plane, where it becomes a bound state with negative real
 energy and with purely imaginary $k$. Here the solution becomes  a  real function with finite number of zeros. The nodes number $n$ counts the zeros of the
 bound state wave function outside  $r=0$.
 
  Antibound solutions belong to poles of $S(k)$ on the negative wing of the imaginary
  $k$-axis. They  diverge without oscillations in the asymptotic region as $r\rightarrow\infty$. 
 The poles of the $S(k)$ lying off the imaginary $k$-axis  on the lower half of the $k$-plane are resonances.
 For a resonance the wave function is complex and the tail of the wave function (in both the
 real and the imaginary parts)  oscillates around  $r$-axis with exponentially growing amplitude as $r\rightarrow\infty$. Both parts have infinite number of zeros. 
 
  It was shown earlier~\cite{[Sa08]} that in the CWS potential the
positions of broad resonances  do depend on the value of $R_{\rm max}$~\cite{[Ra11],[Da12]}, therefore, the
cut-off radius is an important but unphysical parameter of the CWS form in
Eq.~(\ref{vagottWS}). 

\section{Numerical example}

Here we study  the positions of the
poles for  $l=2$,  where we have a centrifugal barrier.     
Among the  resonances we have poles with small imaginary parts. These  narrow resonances appear only if we have a barrier in the potential. 
As example we calculate the neutron  single-particle resonances in
the $^{208}$Pb$+n$ system in the CWS potential used in Ref. \cite{[Cu89]} but without the spin-orbit term.
The values of the potential parameters are: $V_0=44.4$ MeV, $R=r_0 208^{1/3}$, $r_0=1.27$, $a=0.7$. Here we present  only $l=2$ results, which show a typical
example for the $l>0$ cases.

   
\begin{figure}
\resizebox{0.45\textwidth}{!}{\includegraphics{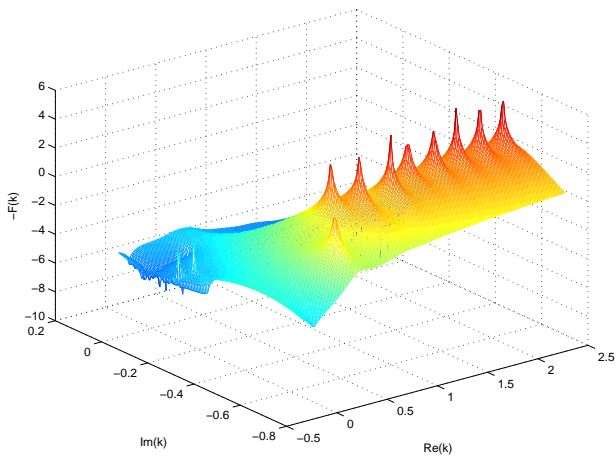}}
\vspace{0.5cm}	   
\caption{Landscape of the  $-F(k)$ surface over the complex wave number domain $k^R\in[-0.1,2.5]$, and $k^I\in[0.1,-0.7]$ for $l=2$, and  $R_{max}=15$.}
\label{modws15}
\end{figure}

In Fig.  \ref{modws15}
 we present the $-F(k)$ surface calculated for $R_{max}=15$ fm for CWS potential.   The peaks of the surface show the approximate pole positions of the
 resonances. They give only a hint for the location of the poles, and help us to give reasonably good starting
  values for finding  the poles of $S(k)$ more accurately, when we calculate
the minima of the quantity $F(k)$ in Eq. (\ref{minima1}) for each pole.
The positions of the three lowest lying poles in the domain $0<|k^R|<0.8$, and $-0.7<k^I<0$
are listed in Table \ref{startl2}. Decaying resonances are denoted
by $d_m$ with $k^R_m>0$, and capturing ones by $c_m$ with $k^R_m<0$.
We assign a sequence number $m$ to the poles as their $|k_m^R|$ value increase.

\begin{table}
\begin{center}
\caption{Sequence numbers $m$, $k_m$ values and node numbers $n$ for the first three lowest lying poles
 for $\ell =2$, $R_{max}=15$ fm, $V_0=44.4$ MeV.}
\begin{tabular}{cccc}
\hline\hline
$~~m~~$ & $~~|k^R_m|~~$ & $~~k^I_m~~$ & $~~n~~$\\
\hline\hline
1&  0.076 &-0.147   &     4\\
2&  0.438     & -0.603  &     3\\  
3&  0.752     & -0.449  &     5 \\
\hline
\end{tabular}
\label{startl2}
\end{center}
\end{table}
Note that the node number in the fourth column does not increase monotonously
with $m$ in the first column. Observe also that the poles with $m=1$ and  $m=2$ are sub-threshold resonances.

In the left corner of the landscape one can see a tiny  pair of poles.
They are the $m=1$ poles in Table \ref{startl2} ($c_1,d_1$). On the right and middle part of the figure we can see a mountain with almost equidistant peaks.

The $d_3$ resonance is a member of this peak-sequence lying closest to the imaginary axis. Below this  $d_3$ pole lies the $d_2$ sub-threshold resonance with considerably larger $|k^I|$ value than the rest of the peaks forming the  mountain.

\begin{figure}
\resizebox{0.45\textwidth}{!}{\includegraphics{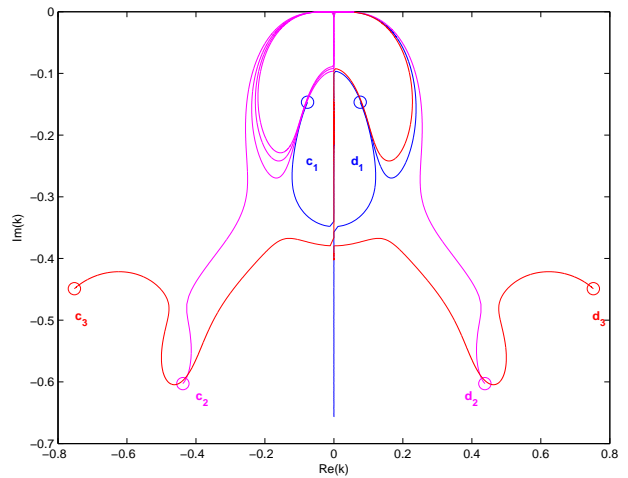}}
\vspace{0.5cm}	   
\caption{Trajectories of the first three poles as the depth of the CWS potential increases. Starting points of the  trajectories at $V_0=44.4$ MeV are denoted by circles. We use $R_{max}=15$ fm.}
\label{CWSL2}
\end{figure}

To understand the dynamics of the poles it is very useful to produce an animation (see Ref. \cite{[anim]}) in which the potential depth is changing by $\Delta V_0$ in each
temporal step and we can see how the landscape $-F(k)$ changes in a given $k$ domain as a function of $V_0$. In the animation we can conveniently  follow the move of the poles as a function of $V_0$.
 In Fig. \ref{CWSL2} we show the trajectories of the poles with starting positions listed in Table \ref{startl2}. We plot their
 trajectories  in a domain in the third and forth quadrants.


\begin{figure}
\resizebox{0.45\textwidth}{!}{\includegraphics{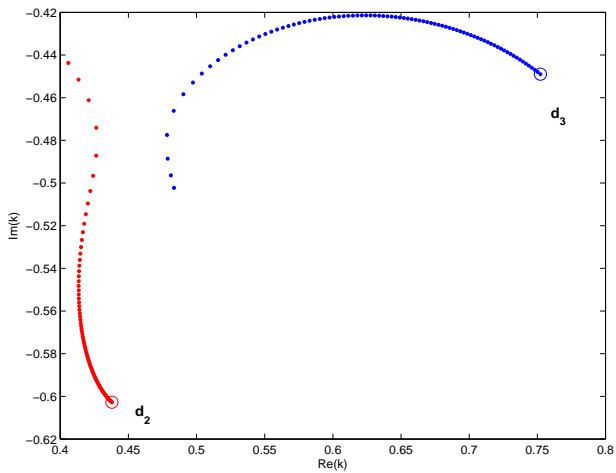}}
\vspace{0.5cm}	   
\caption{Repulsion of resonances $d_2$ and $d_3$  as the depth of the CWS potential increases with steps $\Delta V=0.1$ MeV. Starting points of the  trajectories at $V_0=44.4$ MeV are denoted by circles. Their position is the closest at $V_0=51.5$ MeV.}
\label{taszit}
\end{figure}

By increasing  $V_0$ starting from $V_0=44.4$ MeV the pair $(c_1,d_1)$ in Fig. \ref{CWSL2} moves downward and the pair $(c_2,d_2)$ upward. The third pair $(c_3,d_3)$ moves toward the imaginary axis and interacts with the second pair
 $(c_2,d_2)$. The  repulsive interaction between them is shown in a magnified scale in
 Fig. \ref{taszit}.
The $d_2$ and $d_3$ resonances repel each other when they come close at $V_0=51.5$ MeV. The repulsion changes the direction of the $d_2$ trajectory  slightly towards the imaginary axis, while $d_3$ turns down and move farther from the imaginary axis. 
  The repulsion  
 produces some sort of {\it avoided crossing} of the resonance levels.  For narrow resonances similar phenomenon was observed in axially deformed Woods-Saxon potential
 in Ref.  \cite{[Fe97]}  when the deformation  was changed.
The repulsive interaction is extremely sensitive to the value of $R_{max}$. A small change from $R_{max}=15$ fm to $R_{max}=15.1$ fm reduced the closest distance between the resonances to half of the distance shown in Fig. \ref{taszit}.
 
Increasing the $V_0$ value further
the $(c_2,d_2)$ pair of poles  interacts with the pair $(c_1,d_1)$ and pushes it to
 the imaginary axis at about $k=(0,-0.35)$. After the collision  the pair $(c_2,d_2)$ move upward and the $c_2$ pole collides in the origin with the $d_2$ pole. They form a double pole there. Increasing the strength further the $c_2$ pole becomes to be a bound pole with node number $n=3$.
The other pole what we still call as $d_2$ moves down along the imaginary axis. (In identifying the separating poles from the double pole in the origin, we shifted them a bit from
  the origin by applying a small imaginary potential strength.) At $V_0=66.4$ MeV $d_2$ antibound pole meets the upward moving $c_1$ antibound pole at about $k=(0,-0.1)$ on the imaginary axis. After the collision of the two antibound poles they move off the axis,
 $c_1$ turns left,  $d_2$ turns right  and they become resonances again.
  The $c_1$ and  $d_2$ resonances move  along cucumber like paths symmetrically to the imaginary axis and at the ends of the paths they collide in the origin forming a double pole there. After that the $c_1$ pole becomes to be a bound state pole with node number $n=4$, and  $d_2$ pole starts  moving down again along the imaginary axis as an antibound pole. 
We shall see that the pole $d_2$ plays the role of a {\it bouncer}, since it creates
bound state poles from the double-poles in the origin.

An unexpected feature of the $l=2$ case is that resonances meet with $d_2$ pole twice and they form double pole twice on the imaginary axis. First when the cucumber-like path starts and the second time when they  meet in the origin.

The same pattern is repeated again and again, i.e. the members of the symmetric pair of poles
meet on the imaginary axis well below the origin and
the member coming from the forth quadrant turns down, the other coming from the third quadrant turns up and collides higher with the bouncer ($d_2$) twice.
\section{Summary}
As a summary, we can conclude that for CWS potentials the
$l=2$ pole positions are sensitive to the value of the cut-off radius and their
interaction is extremely sensitive to it.
The distance of the poles in the mountain is regulated by the $R_{max}$ value, as it was observed in Ref. \cite{[Sa08]} for $l=5$.
The larger the $R_{max}$ value is, the closer the peaks are to each other. The close lying poles might interact with each other. When the resonances come close to the imaginary axis they might become antibound states and they might interact with other antibound poles moving along the imaginary axis. 
The interactions between poles distort the pole trajectories.
Such distorted trajectories have been observed earlier in Ref. \cite{[Sa14]} where the starting values of the $l=0$
trajectories were calculated for $^{18}$F. (Starting value of the trajectory is, when
 $V_0$ is very small).
In that paper it was observed that the positions of the starting values of the $l=0$ resonances are governed
by the  $R_{max}$ value.
In Fig. 8. of that reference the  $n=1$ trajectory shows a loop.
Strange circulating shapes were observed for $l=0$ also  in Fig. 1. of Ref. \cite{[Ve12]} for the  $n=1$ pole trajectories of the CWS for different  $R_{max}$  values.

It is reasonable to assume that the repulsive interaction between poles is responsible for all these strange trajectory shapes. Large value of $R_{max}$
pushes the resonance pole too close to each other, where they interact. To reduce the possibility of this interaction we suggest to use a relatively small value for the cut-off radius not much larger than the range of the best fit SV potential.

\section*{Acknowledgement}

Authors are grateful to I. Horny\'ak for valuable discussions.
This work was  supported by the OTKA Grant No. K112962.

Published in Eur. Phys. J. A (2015) {\bf 51}: 76.\\
DOI 10.1140/epja/i2015-15076-1
\end{document}